\documentclass[prapplied,
 twocolumn,
 superscriptaddress,groupedaddress,
 amsmath,amssymb
]{revtex4-2}

\usepackage{times}
\usepackage{color}
\usepackage{graphicx}
\usepackage[dvipsnames]{xcolor}
\usepackage{physics}
\usepackage{bm}
\usepackage{mathtools}
\usepackage{upgreek}
\usepackage[colorlinks=true,allcolors=blue]{hyperref}
\usepackage{footmisc}
\usepackage{makecell}
\usepackage{mathtools}
\usepackage{soul}
\usepackage{siunitx}
\usepackage{xcolor}

\newcommand{\kb}{k_\text{B}}

\begin{document}

\title{Stabilizing and improving qubit coherence by engineering noise spectrum of two-level systems}

\author{Xinyuan You}
\email[xinyuan@fnal.gov]{}
\affiliation{Superconducting Quantum Materials and Systems Center, Fermi National Accelerator Laboratory (FNAL), Batavia, IL 60510, USA}

\author{Ziwen Huang}
\affiliation{Superconducting Quantum Materials and Systems Center, Fermi National Accelerator Laboratory (FNAL), Batavia, IL 60510, USA}

\author{Ugur Alyanak}
\affiliation{Superconducting Quantum Materials and Systems Center, Fermi National Accelerator Laboratory (FNAL), Batavia, IL 60510, USA}

\author{Alexander Romanenko}
\affiliation{Superconducting Quantum Materials and Systems Center, Fermi National Accelerator Laboratory (FNAL), Batavia, IL 60510, USA}

\author{Anna Grassellino}
\affiliation{Superconducting Quantum Materials and Systems Center, Fermi National Accelerator Laboratory (FNAL), Batavia, IL 60510, USA}

\author{Shaojiang Zhu}
\email[szhu26@fnal.gov]{}
\affiliation{Superconducting Quantum Materials and Systems Center, Fermi National Accelerator Laboratory (FNAL), Batavia, IL 60510, USA}

\begin{abstract}

Superconducting circuits are a leading platform for quantum computing. However, their coherence times are still limited and exhibit temporal fluctuations.  
Those phenomena are often attributed to the coupling between qubits and material defects that can be well described as an ensemble of two-level systems (TLSs).
Among them, charge fluctuators inside amorphous oxide layers contribute to both low-frequency $1/f$ charge noise and high-frequency dielectric loss, causing fast qubit dephasing and relaxation. 
Moreover, spectral diffusion from mutual TLS interactions varies the noise amplitude over time, fluctuating the qubit lifetime.
Here, we propose to mitigate those harmful effects by engineering the relevant TLS noise spectral densities.
Specifically, our protocols smooth the high-frequency noise spectrum and suppress the low-frequency noise amplitude via depolarizing and dephasing the TLSs, respectively.
As a result, we predict a drastic stabilization in qubit lifetime and an increase in qubit pure dephasing time. 
Our detailed analysis of feasible experimental implementations shows that the improvement is not compromised by spurious coupling from the applied noise to the qubit.

\end{abstract}
\maketitle

\section{Introduction}\label{sec:intro}
Superconducting circuits hold substantial promise as quantum bits due to their great flexibility in circuit design~\cite{Krantz2019a,Martinis2020,Blais2020,blais2021,Gao2021}.
One main challenge of superconducting qubits is their relatively short coherence times, caused by the uncontrolled interaction with solid state environment~\cite{Ithier2005,Oliver2013,Siddiqi2021,Gyenis2021a}. 
Over the past two decades, the coherence times have been boosted more than five orders of magnitudes, and have now reached the millisecond range~\cite{Gyenis2019,Kalashnikov2019,Zhang2020,Somoroff2021}.
Even with these great improvements, longer coherence times are still required to achieve the fault-tolerant computation~\cite{divincenzo,preskill1998fault}. 

Towards understanding and reducing qubit decoherence, multiple noise channels have been identified~\cite{Ithier2005,Oliver2013,Siddiqi2021,Gyenis2021a}.
One popular model capable of describing several of the dominating noise sources is the two-level systems (TLSs)~\cite{Ashhab2006,Paladino2014,Clemens2019,McRae2020, Murray2021}.
In particular, those charged fluctuators inside the amorphous oxide layers of the circuit device couple to qubit modes via electric-dipole interaction, causing qubit decoherence.
At high frequency, they contribute to dielectric loss, limiting the qubit lifetime $T_1$~\cite{Martinis2005,Wang2015e}. 
At low frequency, they lead to $1/f$ charge noise, degrading the qubit pure dephasing time $T_\phi$~\cite{Astafiev2004,Schlor2019,Tomonaga2021}. 
To protect the qubit from those detrimental effects, several approaches have been taken. 
First, the negative impacts of TLSs can be mitigated through advancement in material science~\cite{Oliver2013,Siddiqi2021,Murray2021}. 
Recently, the lifetime of transmon qubits has been improved several times by solely switching the host material from aluminum/niobium to tantalum~\cite{Place2020,Wang2022}.
Second, the coupling and sensitivity of the qubit to TLSs can be greatly suppressed by exploiting freedom in qubit design~\cite{Gyenis2021a}. 
For example, with significantly increased ratio of Josephson junction energy to charging energy, transmon qubits demonstrate exponential insensitivity to $1/f$ charge noise, and thus long pure dephasing time~\cite{Koch2007,Houck2007}.
Besides optimizing circuit parameters, another useful method to suppress the qubit--TLS coupling is to place coplanar 2D circuits into 3D cavities~\cite{Paik2011}. This strongly reduces the electric field energy stored in the oxide layers, and therefore improves qubit lifetime~\cite{Wang2015e}.

In addition to long coherence times, reliable quantum computation also requires them to be stable over time, cooldowns, devices, etc. 
The latter is attracting more attention recently, with ubiquitous experimental evidence showing that the temporal fluctuation of qubit lifetime becomes stronger as its average value gets longer~\cite{Klimov2018,Burnett2019,Simbierowicz2021,Carroll2021}.
The origin of this temporal fluctuation is often attributed to spectral diffusion, proposed by M{\"u}ller \textit{et al.}~\cite{Muller2015}. 
In this model, fluctuating qubit lifetimes are caused by random shifts in the energies of those close-resonance TLSs, which are further induced by the interaction between TLSs and low-frequency two-level fluctuators (TLFs).
To suppress this fluctuation, Matityahu \textit{et al.}~\cite{Matityahu2021} propose to drive the qubit coherently, such that the TLS bath is not only probed at the qubit frequency, but sampled across a larger range of the spectrum. 
In addition, Zhao \textit{et al.}~\cite{Zhao2022} explore the possibility to tune the qubit away from those high-frequency TLSs via ac Stark shift. 
Since both proposals involve coherently driving the qubit, more complex single- and two-qubit gate schemes are required~\cite{Huang}. 

\begin{figure}[htp]
    \centering
    \includegraphics[width=0.45\textwidth]{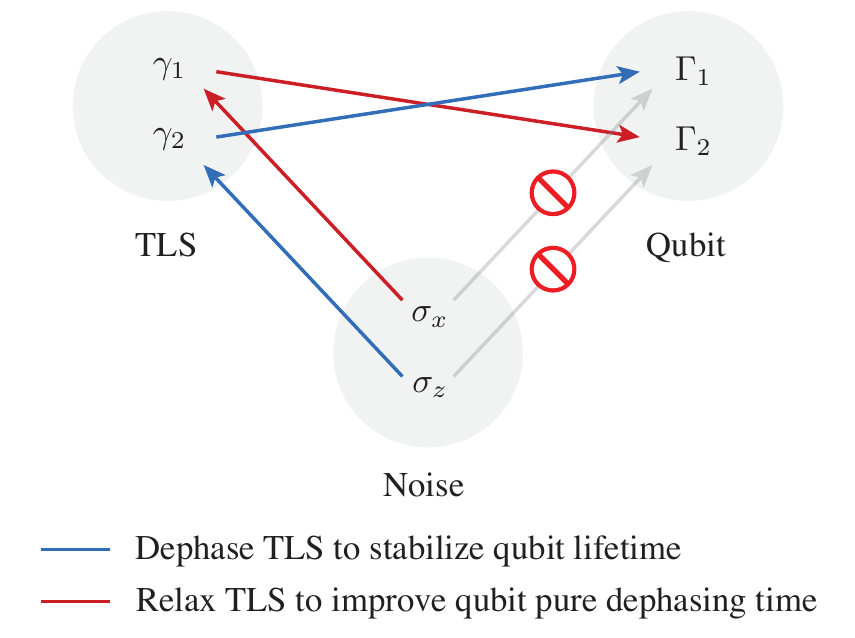}
    \caption{Schematic representation of the protocols. 
    Artificial longitudinal noise is applied to dephase the TLS, which stabilizes the qubit lifetime (blue). 
    Transverse noise is introduced to depolarize the TLS, leading to longer pure dephasing time of the qubit (red). 
    Spurious coupling from the applied noise to the qubit is irrelevant (gray). }
    \label{fig:protocol} 
\end{figure}

Instead of driving the qubit, it is also possible and potentially advantageous to directly address the TLSs~\cite{Burin2013,Khalil2014,Matityahu2019,Yu2021,Burgelman2022,Lisenfeld2022}. 
In this work, we propose to engineer the TLS noise spectrum by implementing artificial noise.
Specifically, we apply longitudinal noise to dephase TLSs, which smooths the high-frequency noise spectrum, and results in stabilization of qubit lifetime. 
Moreover, we utilize transverse noise to depolarize TLSs, which suppress the low-frequency noise amplitude, and leads to an increase of qubit pure dephasing time.
The protocols are summarized schematically in Fig.~\ref{fig:protocol}. 
Compared with the methods that involve driving the qubit, our approach is compatible with standard gate schemes of qubit controls. 

The paper is structured as follows: 
In Sec.~\ref{sec:tls}, we review the standard tunneling model of the TLS, and connect its noise spectral densities to qubit decoherence. 
We then introduce our protocols to improve qubit coherence via engineering TLS noise spectrum in Secs.~\ref{sec:t1} and \ref{sec:t2}, where we show that the qubit lifetime gets stabilized by dephasing TLSs, and its pure dephasing time is improved by depolarizing TLSs. 
We explore feasible experimental realizations in Sec.~\ref{sec:spurious}, and confirm the irrelevance of the spurious coupling from the applied noise to the qubit. 
We share our conclusions in Sec.~\ref{sec:conclusions}, and provide additional details in the subsequent appendix.

\section{TLS as a noise source of qubit decoherence}\label{sec:tls}
In this section, we start with a brief review of the TLS model used in the rest of the work. We then relate the decoherence properties of qubits to noise spectral densities of each single TLS. 
This motivates our protocols developed in later sections. 
We end this section by introducing two different statistics in the presence of an ensemble of TLSs, which are prerequisites to analyze the average and fluctuation of qubit decoherence properties. 

\subsection{Standard tunneling model}
Based on the standard tunneling model~\cite{phillips1981amorphous}, the dynamics of a TLS can be characterized by a double-well potential in some configuration space, with bias energy $\varepsilon$, tunneling amplitude $\Delta$, and thus eigenenergy $\omega_\text{t}=\sqrt{\varepsilon^2+\Delta^2}$. 
With Pauli operators $\{\hat{\Sigma}_i\}$ defined in this configuration basis, the Hamiltonian of an isolated TLS is
\begin{equation}\label{eq:ham_tls}
    \hat{H}_\textrm{TLS} = \dfrac{1}{2} (\varepsilon\, \hat{\Sigma}_z + \Delta\, \hat{\Sigma}_x).
\end{equation}
The microscopic parameters of an individual TLS can be extracted experimentally by probing the spectroscopy of a coupled qubit~\cite{Ashhab2006a,Tian2007,Lupascu2009,Zaretskey2013,Abdurakhimov2021}.
In practice, a TLS couples to some environment, e.g., phonon bath, which leads to its relaxation and dephasing. To describe TLS decoherence, it is convenient to switch to the eigenbasis $\{\hat{\sigma}_i\}$ with the following transformations,
\begin{equation*}
    \hat{\Sigma}_z = \cos\theta \,\hat{\sigma}_z - \sin\theta\, \hat{\sigma}_x, \quad
    \hat{\Sigma}_x = \sin\theta \,\hat{\sigma}_z + \cos\theta \, \hat{\sigma}_x,
\end{equation*}
where the angle $\theta$ follows $\tan \theta = \Delta/\varepsilon$. 
The decoherence dynamics can then be described with a density matrix $\rho(t)$ obeying the Lindblad master equation~\cite{breuer2002theory},
\begin{equation*}
    \dfrac{\mathrm{d}\rho(t)}{\mathrm{d}t}  = -i [\hat{H}_\text{TLS}, \rho(t)] + \gamma_\downarrow\mathcal{D}[\hat{\sigma}_-]\rho + \gamma_\uparrow\mathcal{D}[\hat{\sigma}_+]\rho + \gamma_\phi\mathcal{D}[\hat{\sigma}_z]\rho,
\end{equation*}
with the dissipator defined as $\mathcal{D}[\hat{\mathcal{O}}]\rho = \hat{\mathcal{O}}\rho\hat{\mathcal{O}}^\dag - \{\hat{\mathcal{O}}^\dag\hat{\mathcal{O}},\rho\}/2$.
Taking the phonon bath as the dominant TLS decoherence channel~\cite{phillips1981amorphous}, the relaxation and excitation rates are given by
\begin{equation*}
    \gamma_\downarrow = 2\pi\dfrac{\Delta^2}{\omega_\text{t}^2}
        J(\omega_\text{t}) [n_\text{B}(\omega_\text{t}) +1], \quad \gamma_\uparrow = 2\pi\dfrac{\Delta^2}{\omega_\text{t}^2}
        J(\omega_\text{t}) n_\text{B}(\omega_\text{t}),
\end{equation*}
where $J(\omega)=J_0 \omega^3 e^{-\omega^2/2\omega_\text{D}^2}$ is the typical phonon bath spectral density with Debye cutoff frequency $\omega_\text{D}$, and $n_\text{B}(\omega)$ is the Bose--Einstein distribution. 
To directly measure the decoherence properties of a TLS, microwave pulses are used to initialize the TLS, which time evolution is then mapped onto the population of a coupled qubit for readout~\cite{Lisenfeld2016}. 
Since the parameters $\varepsilon$ and $\Delta$ cover a broad range of frequencies, the lifetime of a TLS varies from nanoseconds to hours~\cite{Constantin2009}. This motivates the idea to use those long-lived TLSs as quantum memory ~\cite{Zagoskin2006,Neeley2008}. 
In contrast, the dephasing rate $\gamma_\phi$ observed in the experiment is usually on the order of MHz~\cite{Lisenfeld2016,Lisenfeld2019}. 
The above master equation thus fully characterizes the dynamics of a single TLS coupled to an environment. 
Next, we study the effects when such a noisy TLS is coupled to a qubit.

\subsection{Noise spectral densities of a single TLS}\label{subsec:psd}
A TLS may couple to a qubit via several ways, resulting in different noise channels~\cite{Clemens2019}. 
For a charged TLS with electric dipole moment, it couples to the electric field of qubit modes, which causes charge noise dephasing~\cite{Astafiev2004,Schlor2019,Tomonaga2021} and dielectric loss relaxation~\cite{Martinis2005,Wang2015e}. 
Similarly, a TLS with magnetic moment may contribute to the total flux in a circuit loop, leading to flux noise~\cite{Yoshihara2006,Bialczak2007,Yan2016a, Kumar2016b,Quintana2017,Braumuller2020}. 
Moreover, a TLS inside the junction oxide layer may block some channels through which Cooper pairs tunnel, and generate critical current noise~\cite{VanHarlingen2004,Constantin2007}. 
Here, we are in particular interested in those charged fluctuators responsible for charge noise and dielectric loss, which are usually the limiting noise sources of many types of superconducting qubits. 
The corresponding electric-dipole interaction between a qubit and a TLS is given by
\begin{equation}
    \hat{H}_\text{qubit--TLS} = \kappa \hat{n} \hat{\Sigma}_z = 
     \kappa \hat{n} (\cos\theta \,\hat{\sigma}_z - \sin\theta\, \hat{\sigma}_x),
\end{equation}
where $\kappa$ denotes the coupling strength, $\hat{n}$ is the qubit charge operator proportional to its electric field, and $\hat{\Sigma}_z$ is related to the TLS electric dipole moment $p\hat{\Sigma}_z$.
From Fermi's golden rule~\cite{Ithier2005}, the relaxation rate $\Gamma_\downarrow$ and excitation rate $\Gamma_\uparrow$ of the qubit are proportional to the relevant TLS noise spectral density $s(\omega)$, evaluating at positive and negative qubit frequencies, i.e.,
\begin{equation}\label{eq:q_t1}
    \Gamma_\downarrow \propto s(\omega_\text{q}), \quad 
    \Gamma_\uparrow \propto s(-\omega_\text{q}).
\end{equation}
In thermal equilibrium, they follow the detailed balance relation $s(-\omega_\text{q}) = s(\omega_\text{q})e^{-\beta\omega_\text{q}}$, with temperature $k_\text{B}T=1/\beta$. 
The qubit dephasing rate is related to the noise spectral density close to zero frequency~\footnote{For spectral density that is singular at zero frequency, a decoherence function has to be involved, see Eq.~\eqref{eq:decoherence_function}},
\begin{equation}\label{eq:q_t2}
    \Gamma_\phi \propto s(0). 
\end{equation}
From Eqs.~\eqref{eq:q_t1} and \eqref{eq:q_t2}, the qubit decoherence dynamics can be fully characterized by the TLS noise spectral density, which is derived in the following. 

The noise spectral density is defined as the Fourier transform of the autocorrelation function of the noise operator~\cite{Clerk2010}, i.e., $\hat{\Sigma}_z$ for the electric-dipole interaction considered here,
\begin{equation}\label{eq:ba_spd}
     s(\omega) = \int_{-\infty}^{+\infty} \mathrm{d}t \,e^{i\omega t} \big[\langle  \hat{\Sigma}_z(t) \hat{\Sigma}_z(0) \rangle - \langle  \hat{\Sigma}_z \rangle^2 \big].
\end{equation}
The notation $\langle \cdots \rangle$ refers to the quantum-mechanical expectation in thermal equilibrium.
In the eigenbasis, Eq.~\eqref{eq:ba_spd} converts to 
\begin{equation}
    s(\omega) = \cos^2 \theta \,s_{zz}(\omega) + \sin^2\theta\, s_{xx} (\omega),
\end{equation}
where $s_{\alpha\alpha}(\omega)$ denotes the spectral density of $\hat{\sigma}_\alpha$.
Note that the cross terms vanish in thermal equilibrium. 
With the Markov approximation~\footnote{Treatment beyond the Markov approximation can be found, e.g., in Ref.~\cite{You2021}}, we apply quantum regression theorem~\cite{breuer2002theory} to derive the relevant noise spectral densities~\cite{shnirman2005low,Constantin2009},
\begin{align}
    \begin{split}
        s_{xx}(\omega) &= \,\dfrac{1-\langle \hat{\sigma}_z \rangle_\text{eq}}{2} \dfrac{2\gamma_2}{(\omega-\omega_\text{t})^2 + \gamma_2^2} \\ 
        &+ \dfrac{1+\langle \hat{\sigma}_z \rangle_\text{eq}}{2} \dfrac{2\gamma_2}{(\omega+\omega_\text{t})^2 + \gamma_2^2},
    \end{split}\label{eq:sxx}\\
    s_{zz}(\omega) &= \,(1-\langle \hat{\sigma}_z \rangle_\text{eq} ^2) \dfrac{2\gamma_1}{\omega^2  + \gamma_1^2}. \label{eq:szz}
\end{align}
Here, $\gamma_1 = \gamma_\downarrow + \gamma_\uparrow$ is the depolarization rate, $\gamma_2 = \gamma_1/2+\gamma_\phi$ the dephasing rate, and $\langle \hat{\sigma}_z\rangle_\text{eq}= (\gamma_\uparrow - \gamma_\downarrow)/(\gamma_\uparrow + \gamma_\downarrow)$ the equilibrium polarization.
Since $s_{xx}(\omega)$ has two peaks at $\omega=\pm\omega_\text{t}$, it mainly contributes to qubit depolarization. 
In contrast, $s_{zz}(\omega)$ is a Lorentzian centered at $\omega=0$, and is responsible for qubit dephasing. 
Following Eqs.~\eqref{eq:q_t1} and \eqref{eq:q_t2}, the qubit depolarization and pure dephasing rates are 
\begin{align}
    \Gamma_1 =&\, \Gamma_\downarrow + \Gamma_\uparrow \approx 
    \kappa^2 M_1 ^2 \sin^2\theta
    \dfrac{2\gamma_2}{\omega_\delta^2 + \gamma_2^2}, \label{eq:t1}\\ 
    \Gamma_\phi =& \kappa^2 M_\phi ^2   \cos^2\theta\,(1-\langle \hat{\sigma}_z \rangle_\text{eq} ^2)/ \gamma_1,\label{eq:t2}
\end{align}
with detuning $\omega_\delta = \omega_\text{q} - \omega_\text{t}$, and matrix elements $M_1=|\langle 0 | \hat{n} | 1\rangle |$, $M_\phi=|\langle 0 | \hat{n} | 0\rangle - \langle 1 | \hat{n} | 1 \rangle |$.
Here, $|0\rangle$ and $|1\rangle$ denote the two logical states of the qubit.

Equations~\eqref{eq:t1} and \eqref{eq:t2} connect the decoherence properties of a qubit ($\Gamma_{1}$, $\Gamma_{\phi}$) to the ones of a TLS ($\gamma_{1}$, $\gamma_{\phi}$), and thus provide intriguing insights on qubit coherence improvement.
First, this relation is asymmetric, in the sense that the lifetime (dephasing time) of the qubit is determined by the dephasing rate (depolarization rate) of the TLS (also illustrated schematically in Fig.~\ref{fig:protocol}). 
Moreover, the decoherence properties of the two subsystems are \textit{negatively} correlated. Specifically, a larger $\gamma_1$ of the TLS results in a smaller $\Gamma_\phi$ of the qubit, and a stronger $\gamma_\phi$ of a close-resonance ($|\omega_\delta|<\gamma_2$) TLS leads to a weaker qubit $\Gamma_1$. 
The above observations motivate our protocols to improve qubit coherence times by engineering TLS noise properties, which will be discussed in detail later in Secs.~\ref{sec:t1} and~\ref{sec:t2}.

\subsection{Statistics of an ensemble of TLSs}\label{subsec:ensemble}
Up to this point, we only focus on a single TLS. In practice, there exists an ensemble of TLSs in a typical qubit device.
The density of avoided level crossings in a qubit spectrum due to coupled TLSs is found to be around $10^2$--$10^3$ $\textrm{GHz}^{-1}\mu\textrm{m}^{-3}$~\cite{Clemens2019}. 
Therefore, a statistical treatment is required to describe the effects of TLSs on a qubit. 
Here, it is crucial to first clarify the nature of the statistics, and to specify the relevant random variables. 

\paragraph{Statistics over TLS realizations}
In the literature, TLSs are observed to have a wide range of bias energies and tunneling amplitudes~\cite{Constantin2009}. 
While the exact values of these parameters are random, experimental evidence indicates that they follow certain probability distributions. 
To explain both the observed $1/f$ charge noise and Ohmic noise simultaneously~\cite{Astafiev2004}, Shnirman \textit{et al.}~\cite{shnirman2005low} propose the joint probability distribution to be $P(\varepsilon,\Delta)\propto \varepsilon/\Delta$. 
Physically, such probability distribution could arise from models like Andreev-level fluctuators~\cite{Faoro2005}.
In comparison, Constantin \textit{et al.}~\cite{Constantin2009} suggest an alternative one, $P(\varepsilon,\Delta)\propto 1/\Delta$, which leads to a white noise behavior in the high-frequency noise spectrum. 
Regardless of the forms of distributions, the specific realization of those parameters for an ensemble of TLSs varies between different device samples and cooldown cycles. 
As a result, qubit coherence times fluctuate according to Eqs.~\eqref{eq:t1} and \eqref{eq:t2}.

\paragraph{Statistics over time}
Recently, many experiments observe that the qubit lifetime fluctuates also within a single cooldown, with a timescale between mileseconds to days~\cite{Klimov2018,Burnett2019,Simbierowicz2021}.
One popular explanation based on spectral diffusion is proposed by M{\"u}ller \textit{et al.}~\cite{Muller2015}.
In this model, the relevant TLSs are sorted into two categories based on their energy scales. 
Those in close resonance with the qubit are denoted as high-energy (coherent) TLSs, dominating qubit lifetime. 
The role of the other low-energy (compared to temperature) TLSs, also known as two-level fluctuators (TLFs), is to fluctuate the frequencies of those high-energy TLSs via TLS--TLF interaction. 
Since the qubit lifetime depends sensitively on its detuning with those high-energy TLSs [see Eqs.~\eqref{eq:t1}], this explains the observed instability in qubit lifetime. 
Here, the relevant random variables responsible for the temporal fluctuation are the states of these TLFs, which can be described by the associated polarization $\sigma_z=\pm 1$.

\section{Stabilizing qubit depolarization time by dephasing TLSs}\label{sec:t1}
In the previous section, we suggest that the qubit decoherence due to a single TLS can be suppressed by making the TLS noisy. 
In the following two sections, we elaborate this idea and extend it to the realistic case of an ensemble of TLS with various statistics discussed above. 
We focus on qubit lifetime here and leave the discussion on dephasing time to Sec.~\ref{sec:t2}. 
To benchmark the improvement, we first characterize the average and variance of qubit lifetime without implementing our protocol. 

\subsection{Statistics of qubit lifetime over TLS realizations}
We start by deriving the statistics of qubit depolarization rate over TLS realizations. 
Following Refs.~\cite{shnirman2005low,Constantin2009}, the joint probability distribution of TLS parameters is
\begin{equation}\label{eq:dist}
    P(\varepsilon,\Delta) = \mathcal{N}(\alpha) \mathcal{B}(\varepsilon,\Delta) \dfrac{\varepsilon^\alpha}{\Delta},
\end{equation}
where $\alpha\in \{0,1\}$ accounts for the two distributions discussed in Sec.~\ref{subsec:ensemble}. The boundary function $\mathcal{B}(\varepsilon,\Delta)$ equals unity for $\varepsilon_\text{m}< \varepsilon < \varepsilon_\text{M}$ and $\Delta_\text{m} < \Delta < \Delta_\text{M}$, and vanishes otherwise. 
The set of the parameters is taken from Ref.~\cite{Constantin2009}, with $\varepsilon_\text{m}/\kb = 0$, $\varepsilon_\text{M}/\kb = 4\, \textrm{K}$, $\Delta_\text{m}/\kb = 2\, \mu\textrm{K}$, and $\Delta_\text{M}/\kb = 4\,\textrm{K}$. The normalization factor is given by
\begin{equation}
    \mathcal{N}(\alpha)^{-1}=
       \left(\dfrac{\varepsilon_\textrm{M}+\varepsilon_\textrm{m}}{2}\right)^\alpha(\varepsilon_\textrm{M}-\varepsilon_\textrm{m})\log\left( \dfrac{\Delta_\textrm{M}}{\Delta_\textrm{m}}\right).
\end{equation}
The average qubit depolarization rate is then,
\begin{equation}\label{eq:avg_real}
    \langle \Gamma_1 \rangle_\text{ens} = \iint_{\mathbb{R}^2} \mathrm{d}\varepsilon\, \mathrm{d}\Delta\, P(\varepsilon,\Delta)\, \Gamma_1,
\end{equation}
with $\Gamma_1$ taken from Eq.~\eqref{eq:t1}.
To obtain analytical results, it is useful to perform change of variables from $(\varepsilon,\Delta)$ to $(\omega_\text{t},\gamma_2)$, with the TLS dephasing rate,
\begin{equation}\label{eq:gamma2}
    \gamma_2 = \pi J_0\Delta^2 \omega_\text{t}\coth\left(\beta\omega_\text{t}/2\right) + \gamma_\phi.
\end{equation}
Here the pure dephasing rate $\gamma_\phi=10\,\text{MHz}$ is assumed to be a constant for simplicity~\cite{Lisenfeld2016,Lisenfeld2019}. With some algebra (see details in Appendix~\ref{app:details}), Eq.~\eqref{eq:avg_real} equals
\begin{equation}\label{eq:avg_ens}
     \langle \Gamma_1 \rangle_\text{ens} \approx 
     \begin{cases}
      \kappa^2 M_1 ^2\mathcal{N}\pi\omega_\text{q} , &\alpha = 1 \\ \\
      2\kappa^2 M_1 ^2\mathcal{N}\pi, &\alpha = 0
    \end{cases}.
\end{equation}
The results exhibit the expected Ohmic and white noise behavior for $\alpha=1$ and $\alpha=0$, respectively. 
We note that both expressions do not depend on TLS dephasing rate. This seems to be inconsistent with our observation in Sec.~\ref{subsec:psd}, where the qubit depolarization rate decreases with a larger dephasing rate of a single close-resonance TLS.
The contradiction can be resolved by noticing that  Eq.~\eqref{eq:avg_ens} is derived for an ensemble average of TLSs, which also includes those TLSs that are far-detuned from the qubit ($|\omega_\delta|>\gamma_2$). 
For those TLSs, increasing their dephasing rates enhances the qubit depolarization rate. 
Therefore, on average, the qubit depolarization rate is insensitive to the TLS dephasing rate. 
However, as we will show in the following, this is not the case for the fluctuation of qubit lifetime. 

\begin{figure}
    \centering
    \includegraphics[width=0.45\textwidth]{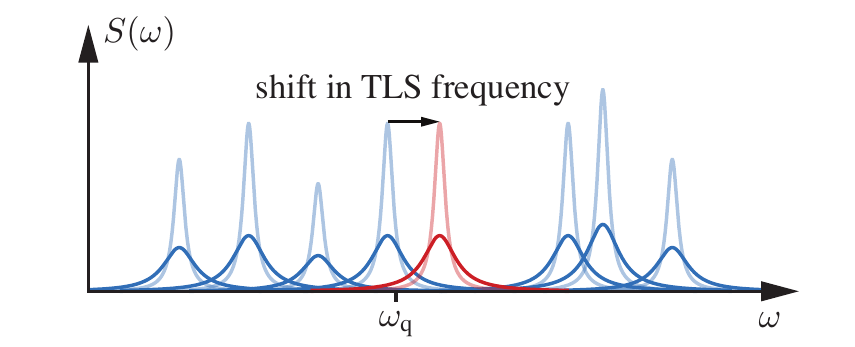}
    \caption{Schematic representation of the noise spectral density of an ensemble of TLSs. 
    In the presence of a random shift in a TLS frequency, the spectral density probed by the qubit at $\omega_\text{q}$ fluctuates less for flattened peaks, corresponding to larger TLS linewidth.}
    \label{fig:fluctuation} 
\end{figure}

To derive the variance of qubit lifetime, we calculate 
\begin{equation}\label{eq:var_ens_org}
    \textrm{Var}_\text{ens}(\Gamma_1)\approx \langle \Gamma_1^2 \rangle_\text{ens} = \iint_{\mathbb{R}^2} \mathrm{d}\varepsilon\, \mathrm{d}\Delta\, P(\varepsilon,\Delta) \Gamma_1^2.
\end{equation}
With details in Appendix~\ref{app:details}, the above expression equals
\begin{equation}\label{eq:var_ens}
     \textrm{Var}_\text{ens}(\Gamma_1) \approx 
     \begin{cases}
      \kappa^4 M_1 ^4\mathcal{N}\pi\omega_\text{q}/2\gamma_\phi , &\alpha = 1 \\ \\
      4\kappa^4 M_1 ^4\mathcal{N}\pi/3\gamma_\phi, &\alpha = 0
    \end{cases}.
\end{equation}
Unlike the average of qubit lifetime, its variance does depend on TLS dephasing rate, and is inversely proportional to that.
Intuitively, this can be understood as follows. 
The fluctuation of qubit lifetime is related to the comb-shape TLS spectral density (each peak corresponds to a TLS), shown schematically in Fig.~\ref{fig:fluctuation}. 
The noise spectral density probed by the qubit $S(\omega_\text{q})$ changes significantly in the case of random shifts in frequencies of some close-resonance TLSs. 
With larger TLS dephasing rate, those peaks are flattened, which suppresses the fluctuation in qubit lifetime.

Note that the above results [Eqs.~\eqref{eq:avg_ens} and \eqref{eq:var_ens}] are the averaged contribution from a single TLS. 
The total depolarization rate of the qubit $\Gamma_{1,\text{tot}}$ sums over numerous TLSs (with number $N_\text{TLS}$). 
Under the assumption that the TLSs are independent and identical (i.e., they obey the same distribution), $\Gamma_{1,\text{tot}}$ follows a normal distribution by central limit theorem. 
Its average and variance are given by
\begin{equation*}
    \langle \Gamma_{1,\text{tot}} \rangle = N_\text{TLS} \langle \Gamma_1\rangle, \quad 
    \text{Var}(\Gamma_{1,\text{tot}}) = N_\text{TLS}\text{Var}(\Gamma_1).
\end{equation*}
Since most experimental data is expressed in terms of $T_1$, it is also useful to derive the statics of $T_{1,\text{tot}}=1/\Gamma_{1,\text{tot}}$, 
\begin{equation*}
    \langle T_{1,\text{tot}} \rangle = 1/\langle \Gamma_{1,\text{tot}} \rangle, \quad
    \text{Var}(T_{1,\text{tot}}) = \text{Var}(\Gamma_{1,\text{tot}}) / \langle \Gamma_{1,\text{tot}} \rangle^4.
\end{equation*}

\subsection{Statistics of qubit lifetime over time}
The analysis above addresses the fluctuation of qubit lifetime over different cooldown cycles and device samples.
However, qubit $T_1$ also fluctuates temporarily during a single cooldown~\cite{Klimov2018,Burnett2019,Simbierowicz2021}. 
Since the mechanisms of the fluctuations are different, the relevant statistics of qubit lifetime may also change.
In the following, we derive the average and variance of qubit lifetime regarding the temporal fluctuation. 

Based on the spectral diffusion model~\cite{Muller2015}, this temporal fluctuation arises from the TLS--TLF interaction, 
\begin{equation}\label{eq:tls_tlf}
    \hat{\mathcal{H}}_{\text{TLS--TLF}} = \dfrac{1}{2}\sum_{i} g_{i} \hat{\sigma}_{z}\hat{\sigma}_{z,i},
\end{equation}
where the subscript $i$ indexes TLFs coupled to the studied TLS (with no subscript), and $g_i$ denotes their interaction strength. 
Since the energy of those TLFs is much smaller than temperature, they can be treated classically, i.e., $\hat{\sigma}_{z,i}\to \sigma_{z,i}$.
The average and variance of these random variables are 
\begin{equation*}
    \langle \sigma_{z,i}\rangle = -\tanh\left(\beta \omega_{\text{t},i}/2\right), \,\,
    \textrm{Var}(\sigma_{z,i}) = 1 - \tanh^2\left(\beta \omega_{\text{t},i}/2\right). 
\end{equation*}
Due to the TLS--TLF interaction, the TLS energy depends on the states of those coupled TLFs,
\begin{equation}
    \omega_{\text{t}} = \omega_{\text{t}}^{(0)} + \sum_{i} g_{i}\sigma_{z,i}, 
\end{equation}
with $\omega_{\text{t}}^{(0)}$ the unperturbed TLS frequency. 
For weak coupling, we can Taylor expand the qubit depolarization rate in Eq.~\eqref{eq:t1}, and find its first-order dependence on the coupling to TLFs,
\begin{equation}\label{eq:gamma_1_tlf}
    \Gamma_{1} = \Gamma_{1}^{(0)} + \Gamma_{1}^{(1)}\sum_{i} g_{i}\sigma_{z,i}. 
\end{equation}
The two coefficients are defined as follows,
\begin{align}
    \begin{split}
        \Gamma_{1}^{(0)} &= \kappa^2 M_1 ^2\sin^2 \theta \dfrac{2\gamma_{2}}{\gamma_{2}^2 + \omega_\delta^2}, \\
    \Gamma_{1}^{(1)} &= \kappa^2 M_1 ^2\sin^2 \theta \dfrac{4\gamma_{2}\omega_\delta}{\left(\gamma_{2}^2 + \omega_\delta^2\right)^2}.
    \end{split}
\end{align}
Then, we take the average of Eq.~\eqref{eq:gamma_1_tlf} regarding the states of those TLFs, 
\begin{equation}\label{eq:avg_sd}
    \langle \Gamma_{1} \rangle_\text{spd}  =  \Gamma_{1}^{(0)}  + \Gamma_{1}^{(1)}\sum_{ i} g_{i}\langle \sigma_{z,i}\rangle  = \Gamma_{1}^{(0)}.
\end{equation}
In deriving the second equality, we assume the distribution of $g_i$ is symmetric about zero. 
As a result, Eq.~\eqref{eq:avg_sd} shows that there is no net effect of spectral diffusion on the average qubit lifetime.

\begin{figure}
    \centering
    \includegraphics[width=0.45\textwidth]{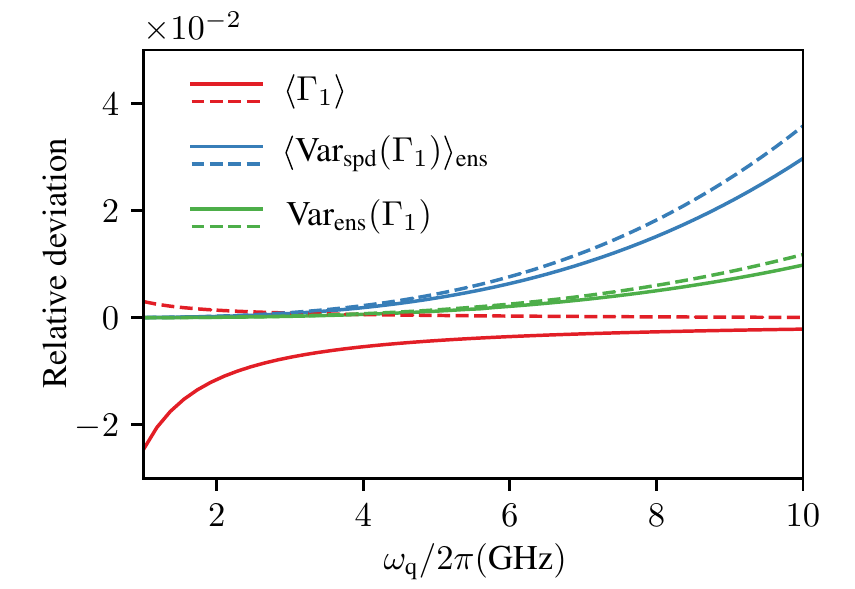}
    \caption{Relative deviations between the approximated analytical expressions in Eq.~\eqref{eq:avg_ens}, \eqref{eq:var_ens}, and \eqref{eq:var_spd}, and the numerically obtained results, as a function of the qubit frequency. 
    The solid and dashed lines correspond to the two joint probability distributions with $\alpha=1$ and $\alpha=0$, respectively. }
    \label{fig:analytics} 
\end{figure}

Next, we derive the variance of qubit lifetime. With the assumption that the TLFs are uncorrelated, we obtain
\begin{equation}\label{eq:var_sd}
    \textrm{Var}_\text{spd} (\Gamma_1) = \langle \Gamma_{1}^2 \rangle_\text{spd} - \langle \Gamma_{1} \rangle_\text{spd}^2
    = \left(\Gamma_{1}^{(1)}\right)^2 G^2,
\end{equation}
where the quantity $G^2=\sum_{i}  g_{i}^2 (1-\tanh^2{\beta \omega_{\text{t},i}/2})$ manifests the fluctuation of TLFs.
Note that both Eqs.~\eqref{eq:avg_sd} and~\eqref{eq:var_sd} are derived for a specific TLS. 
To account for the existence of multiple TLSs, we further take an ensemble average of Eq.~\eqref{eq:var_sd} with respect to TLSs realizations~\footnote{Note that we take the ensemble average of the variance instead of its standard deviation, since $\Gamma_1$ of each TLS follows a normal distribution.}.
\begin{equation}\label{eq:int_spd}
    \langle \textrm{Var}_\text{spd} (\Gamma_1) \rangle _\text{ens} =  \iint_{\mathbb{R}^2} \mathrm{d}\varepsilon\, \mathrm{d}\Delta\, P(\varepsilon,\Delta) \left(\Gamma_{1}^{(1)}\right)^2 G^2.
\end{equation}
For simplicity, we assume $G^2$ to be the same for all TLSs. 
With details in Appendix~\ref{app:details}, we derive
\begin{equation}\label{eq:var_spd}
     \langle \textrm{Var}_\text{spd} (\Gamma_1) \rangle _\text{ens} \approx 
     \begin{cases}
      \kappa^4 G^2 M_1 ^4\mathcal{N}\pi\omega_\text{q}/4\gamma_\phi^3 , &\alpha = 1 \\ \\
      2\kappa^4 G^2 M_1 ^4\mathcal{N}\pi/3\gamma_\phi^3, &\alpha = 0
    \end{cases}.
\end{equation}
Compared with $\textrm{Var}_\text{ens} (\Gamma_1)$ in Eq.~\eqref{eq:var_ens}, $\langle \textrm{Var}_\text{spd} (\Gamma_1) \rangle _\text{ens}$ also depends on TLS dephasing rate, but more sensitively with an inverse cubic dependence.  
This indicates a significant suppression of temporal fluctuation in qubit lifetime by increasing TLS dephasing rate.
To verify the analytical expressions in Eqs.~\eqref{eq:avg_ens}, \eqref{eq:var_ens}, and \eqref{eq:var_spd}, we compare them with the results obtained by numerical integration. 
The relative deviations as a function of the qubit frequency are shown in Fig.~\ref{fig:analytics}, which shows decent agreement with each other.

\subsection{Stabilization of qubit lifetime}
The results derived above suggest that we can stabilize qubit lifetime by increasing the TLS dephasing rate. 
In the following, we describe in detail our stabilization protocol, and quantify the potential improvement. 
In general, a TLS responds to external coupling through its bias energy~\cite{Clemens2019},
\begin{equation}
    \varepsilon = \varepsilon_0 + 2 \mathbf{d}\cdot \mathbf{F},
\end{equation}
with $\varepsilon_0$ the intrinsic TLS bias energy. The external field $\mathbf{F}$ can be an electric field $\mathbf{E}$, or a mechanical strain field $\mathbf{S}$ (not to be confused with the spectral density $S$), and $\mathbf{d}$ is the corresponding electric dipole $\mathbf{d}_\text{e}$ or strain tensor $\mathbf{d}_\text{s}$ of the TLS. 
For a typical TLS, the two coupling strengths are on the orders of  $|\mathbf{d}_\text{s}|\sim 1\text{eV}$ and $|\mathbf{d}_\text{e}|\sim 1\text{D}$, 4respectively~\cite{Clemens2019}. 
Recently, static tuning of TLS frequency has been realized experimentally via both dc electric and strain fields~\cite{Grabovskij2012a,Lisenfeld2015a,Lisenfeld2016,Sarabi2016,Brehm2017d,Lisenfeld2019,Matityahu2019,Bilmes2020a,Bilmes2021}. 
Here, the technique useful for our purpose is instead to apply ac signals, or more precisely, noisy signals. 
The interaction Hamiltonian is given by
\begin{equation}\label{eq:pert}
    \hat{H}_\text{int} = \mathbf{d}\cdot \mathbf{F}\cos\theta\,\hat{\sigma}_z + \mathbf{d}\cdot \mathbf{F}\sin\theta\,\hat{\sigma}_x.
\end{equation}
Note that here the field $\mathbf{F}$ can either be a quantum operator or a classical variable. 
From Fermi's golden rule, $\hat{H}_\text{int}$ results in additional TLS depolarization and pure dephasing,
\begin{gather}
    \delta\gamma_{1} = \delta\gamma_\downarrow + \delta\gamma_\uparrow=  \left( d \cos{\alpha}\sin{\theta} \right)^2 \left[S_\text{add}(\omega_\text{t}) + S_\text{add}(-\omega_\text{t})\right], \nonumber\\
    \delta\gamma_{\phi} = 2\left( d \cos{\alpha}\cos{\theta} \right)^2 S_\text{add}(0)\label{eq:add_decoherence},
\end{gather}
where $S_\text{add}(\omega)$ is the noise spectral density of the applied field $\mathbf{F}$, with an angle $\alpha$ relative to the dipole $\mathbf{d}$. 
Depending on the effective temperature of the noise, characterized by $\kb T = \omega/\log[S_\text{add}(\omega)/S_\text{add}(-\omega)]$, the equilibrium polarization of the TLS may also change, 
\begin{equation}\label{eq:polarization}
    \langle \hat{\sigma}_z \rangle' = \dfrac{ (\gamma_\uparrow+\delta\gamma_\uparrow) - (\gamma_\downarrow+\delta\gamma_\downarrow) }{(\gamma_\uparrow+\delta\gamma_\uparrow) + (\gamma_\downarrow+\delta\gamma_\downarrow)}. 
\end{equation}

To quantify the stabilization of qubit lifetime, we reevaluate the average [Eq.~\eqref{eq:avg_real}] and variance [Eqs.~\eqref{eq:var_ens} and~\eqref{eq:var_sd}] of qubit $\Gamma_1$, with TLS dephasing rate $\gamma_2$ modified from Eq.~\eqref{eq:gamma2},
\begin{equation}
    \gamma_2' = \pi J_0\Delta^2 \omega_\text{t}\coth\left(\beta\omega_\text{t}/2\right) + \gamma_\phi + \delta\gamma_1/2 + \delta\gamma_\phi. 
\end{equation}
Note that the quantity $G^2$ in Eq.~\eqref{eq:var_sd} changes as well due to the modified $\langle \hat{\sigma}_z \rangle'$. 
For the purpose of dephasing TLSs, it is appropriate to apply a low-frequency noise. 
Specifically, we take a classical white noise $S_\text{add}(\omega)=S_\text{add}$, with a high-frequency cutoff $\omega_\text{c}$ on the same order of the TLS dephasing rate. 
Note that as we increase the amplitude of the noise, TLS dephasing rate also grows. Thus, we need to adjust the cutoff frequency accordingly.

\begin{figure}
    \centering
    \includegraphics[width=0.45\textwidth]{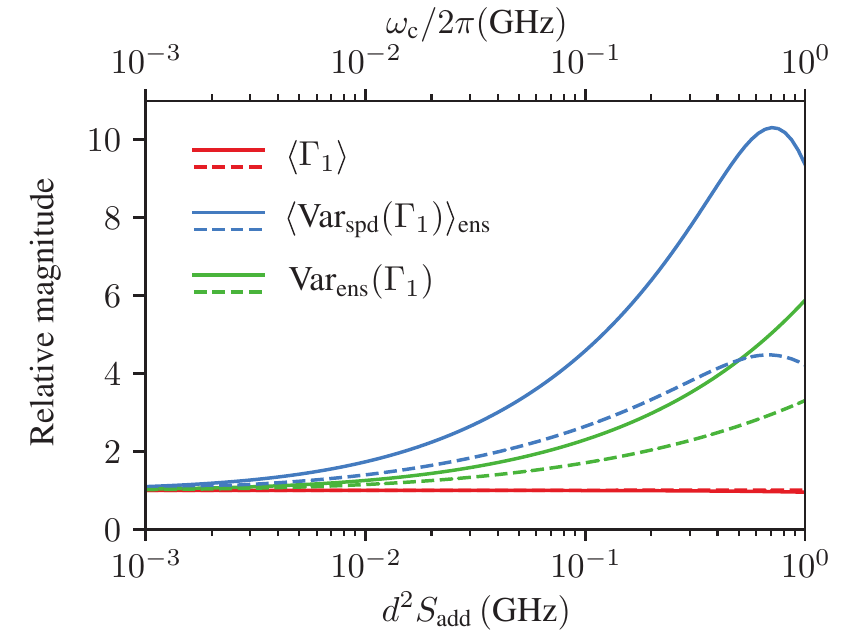}
    \caption{Relative magnitudes of the average and variances of qubit depolarization rates between the cases when the protocol is turned off and on. 
    The lower horizontal axis denotes the strength of the noise, and the upper one shows its cutoff frequency. 
    The solid and dashed lines correspond to two joint probability distributions with $\alpha=1$ and $\alpha=0$, respectively. [Parameters: $\omega_\text{q}/2\pi$ = 5.7 GHz, $T$ = 20 mK, $J_0$ = 0.047 ps$^2$.] }
    \label{fig:stabilization} 
\end{figure}

In Fig.~\ref{fig:stabilization}, we plot the stabilization of qubit lifetime as a function of the strength of the applied noise. 
The shown results are the relative magnitudes of the variances of qubit depolarization rates between the cases when the protocol is turned off and on. 
Here, the noise strength is denoted by $d^2 S_\text{add}$, which is proportional to the additional TLS dephasing rate. 
Alternatively, we can express the noise strength via its standard deviation $\sqrt{S_\textrm{add}\omega_\text{c}/\pi}$. For the largest noise amplitude taken in Fig.~\ref{fig:stabilization}, $d^2S_\text{add}=1$ GHz, it corresponds to a standard deviation of strain field about $10^{-6}$ (e.g., 10-nm deformation for a 10-mm chip) and electric field around $10$ kV/m. 
In addition to the amplitude of the noise, we supplement the amplitude dependent cutoff frequency $\omega_\text{c}$. 
Clearly, both variances $\langle\text{Var}_\text{spd}(\Gamma_1)\rangle_\text{ens}$ and $\text{Var}_\text{ens}(\Gamma_1)$ are suppressed with our protocol, with the latter demonstrating 10 times better stabilization. 
While $\text{Var}_\text{ens}(\Gamma_1)$ exhibits a monotonic suppression, $\langle\text{Var}_\text{spd}(\Gamma_1)\rangle_\text{ens}$ has a minimum regarding the noise amplitude (or maximum for the relative magnitude shown in Fig.~\ref{fig:stabilization}). 
This is because the effective temperature of TLFs are raised by the applied classical noise, causing TLFs to switch more likely (i.e., larger $G^2$)~\footnote{This will not be the case in terms of a quantum noise with identical effective temperature with the TLS bath.}. 
Therefore, even the change in qubit lifetime during a TLF switch is suppressed for stronger applied noise, the frequency of the switch is enhanced. 
These two competing factors result in the local extremum of $\langle\text{Var}_\text{spd}(\Gamma_1)\rangle_\text{ens}$ observed in Fig.~\ref{fig:stabilization}. 
We also notice that the average depolarization rate is almost independent of the applied noise:
With $\alpha=1$, there is a minute 5\% increase of $\langle \Gamma_1 \rangle$ for a 10$\times$ suppression of $\langle\text{Var}_\text{spd}(\Gamma_1)\rangle_\text{ens}$. 

\section{Improving qubit pure dephasing time by depolarizing TLSs}\label{sec:t2}
In Sec.~\ref{sec:t1}, we introduced our protocol to stabilize the qubit lifetime by dephasing TLSs.
Here, we instead explore ways to enhance the qubit pure dephasing time.
Early in Eq.~\eqref{eq:t2}, we derive the dephasing time of the qubit, when it couples to a single TLS.
The generalization to the case of an ensemble of TLSs requires special treatment, since the relevant noise spectral density here is singular around zero frequency (as will be shown in the following).
With the usual assumption of Gaussian noise, the qubit dephasing (free induction) is described by~\cite{Ithier2005},
\begin{equation}\label{eq:decoherence_function}
    \langle e^{i\delta\phi(t)} \rangle = \exp \left[-\dfrac{t^2M^2}{2} \int_{-\infty}^{\infty}
    \dfrac{\mathrm{d}\omega}{2\pi}\,S_\text{low}(\omega)\, \text{sinc}^2\dfrac{\omega t}{2}\right],
\end{equation}
with $M$ the relevant matrix element, and $S_\text{low}(\omega)$ the low-frequency noise spectral density. 
Unlike Eq.~\eqref{eq:t2}, where only the zero frequency noise spectral density is probed, here in Eq.~\eqref{eq:decoherence_function}, the qubit dephasing depends on a broad range of frequencies controlled by the filter function.
In the remaining of the section, we first derive the detailed structure of $S_\text{low}(\omega)$ of an ensemble of TLSs. 
Then, we demonstrate the protocol to suppress $S_\text{low}(\omega)$ by depolarizing TLSs, which leads to longer qubit pure dephasing time. 

\subsection{White and $1/f$ noise from an ensemble of TLSs}
\begin{figure}
    \centering
    \includegraphics[width=0.45\textwidth]{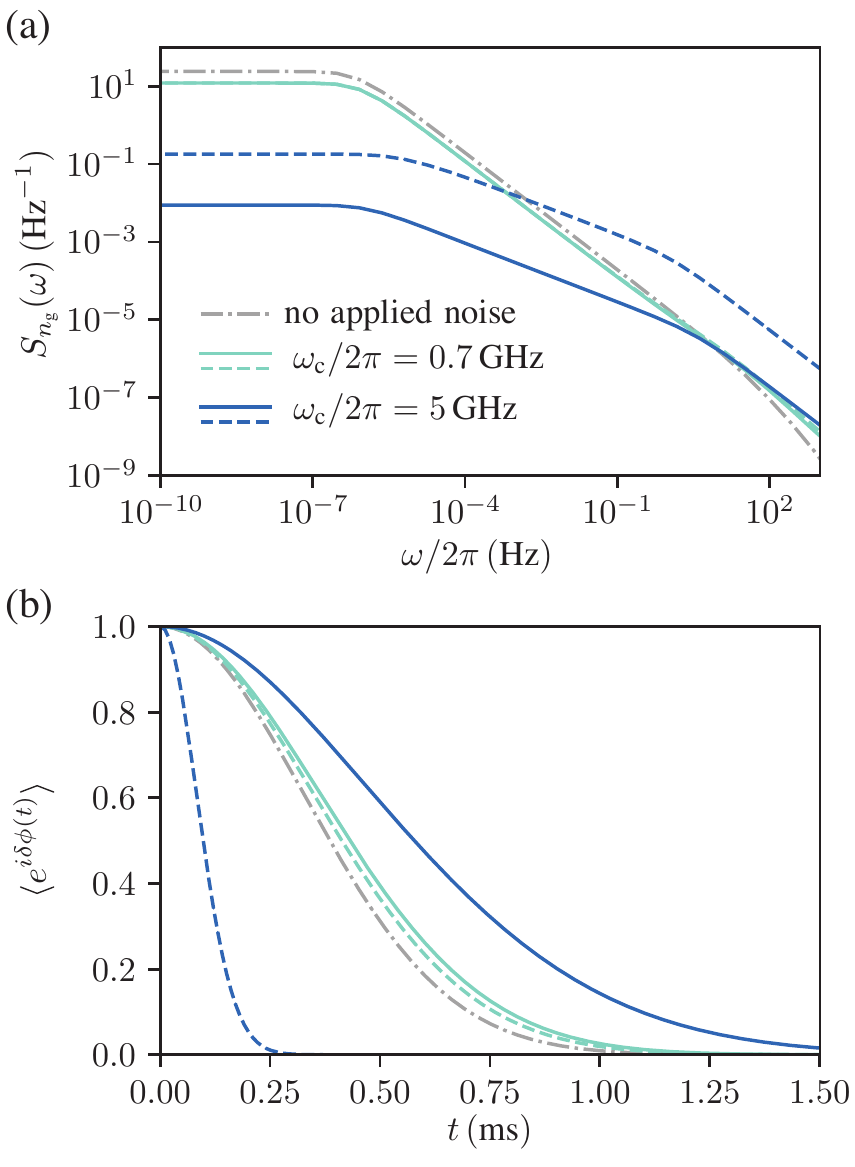}
    \caption{
    (a) Low-frequency noise spectral densities of an ensemble of TLSs. 
    Different colors represent the cutoff frequencies of the applied noise. 
    Solid and dashed lines distinguish quantum and classical noise, respectively. 
    (b) Transmon qubit dephasing dynamics according to the noise spectral densities shown in (a). [Parameters: $d^2 \eta = 1$, $E_\text{C}/h=0.3$ GHz, and $E_\text{J}/h=15$ GHz.] }
    \label{fig:coherence} 
\end{figure}

The low-frequency noise spectral density of an ensemble of TLSs is obtained by taking an average of $s_{zz}(\omega)$ in Eq.~\eqref{eq:szz},
\begin{equation}\label{eq:s_low_orig}
    S_\text{low}(\omega) = N_\text{TLS}\iint_{\mathbb{R}^2}\mathrm{d}\varepsilon\,\mathrm{d}\Delta\, P(\varepsilon,\Delta)\cos^2\theta s_{zz}(\omega).
\end{equation}
Since the qualitative behavior of $S_\text{low}(\omega)$ is identical with the joint probability distribution being $\alpha=0\,\text{and}\,1$~\cite{shnirman2005low,You2021}, we take $P(\varepsilon,\Delta)\propto \varepsilon/\Delta$ (i.e., $\alpha=1$) for simplicity. 
To evaluate Eq.~\eqref{eq:s_low_orig}, we apply variable transformation from $(\varepsilon,\Delta)$ to $(\omega_\text{t},\gamma_1)$,
with the TLS depolarization rate,
\begin{equation}
    \gamma_1 = 2\pi J_0 \Delta^2\omega_\text{t} \coth\left(\beta\omega_\text{t}/2 \right).
\end{equation}
With appropriate approximations (see details in Appendix~\ref{app:details}), we derive analytical expressions of the spectral density for different frequency ranges. 
Of particular interest here is the emergence of white and $1/f$ noise at low frequency~\footnote{The crossover from $1/f$ to a $1/f^2$ noise is discussed in Ref.~\cite{You2021}.},
\begin{equation}\label{eq:s_low}
    S_\text{low}(\omega) = 
    \begin{cases}
       \dfrac{N_\text{TLS}\mathcal{N}\kb T}{ 2\pi J_0 \Delta_\text{m}^2}, &\omega \ll \omega_\text{ir} \\ \\
       \dfrac{2\log(2)\pi N_\text{TLS}\mathcal{N}(\kb T)^2}{\omega}, &\omega \gg \omega_\text{ir}
    \end{cases},
\end{equation}
with the infrared cutoff frequency for $1/f$ noise, $\omega_\text{ir}=4\pi^2\log(2)J_0 \kb T\Delta_\text{m}^2$. 
This is further confirmed by numerical integration of Eq.~\eqref{eq:s_low_orig}, shown as the gray line in Fig.~\ref{fig:coherence}(a). 
Note that we have converted $S_\text{low}(\omega)$  to offset charge noise following $S_{n_\text{g}}(\omega)=(d_\text{e}/eL)^2 S_\text{low}(\omega)$, with $L$ a sample-dependent length scale.
The number of TLS is chosen to match  $S_{n_\text{g}}(\omega)=2\pi A^2/\omega$, with a realistic amplitude $A=10^{-3}$~\cite{Tomonaga2021}. 
Plugging the calculated $S_{n_\text{g}}(\omega)$ into Eq.~\eqref{eq:decoherence_function}, we plot the qubit pure dephasing dynamics in Fig.~\ref{fig:coherence}(b), which shows a typical Gaussian decay profile. 

\subsection{Improvement of qubit pure dephasing time}
The qubit dephasing rate derived in Eq.~\eqref{eq:t2} reveals $\Gamma_\phi\propto 1/\gamma_1$, which indicates that we can improve qubit pure dephasing time by depolarizing the TLSs. 
However, $\Gamma_\phi$ also depends on a prefactor $1-\langle \hat{\sigma}_z\rangle^2$, which is related to the fluctuation of TLSs.  
When TLSs undergo additional depolarization, $\langle \hat{\sigma}_z\rangle$ may deviate from $-\tanh{\beta\omega_\text{t}/2}$, depending on whether the nature of the noise is quantum or classical [see Eq.~\eqref{eq:polarization}].
Therefore, it is necessary to quantify the improvement of qubit dephasing time for both types of noise respectively. 
\paragraph{Quantum noise}
In order to depolarize TLSs, we apply high-frequency noise with an Ohmic noise spectrum $S_\text{add}(\omega>0) =  \theta(\omega_\text{c}-\omega)\,\eta\, \omega$.
In the case of quantum noise with the same effective temperature of TLSs, $S_{n_\text{g}}(\omega)$ is modified due to the additional decay of those TLSs with frequencies lower than the cutoff frequency $\omega_\text{c}$,
\begin{equation}\label{eq:add_g1}
    \gamma_1' = \gamma_1 + \delta\gamma_1.
\end{equation}
When $\omega_\text{c}\gg \kb T$, most of the TLSs contributing to qubit dephasing are affected, resulting in a suppressed white noise spectral density (see details in Appendix~\ref{app:details}),
\begin{equation}\label{eq:s_low_add}
    S_\text{white}' = \dfrac{0.58N_\text{TLS}\mathcal{N}(2k_\text{B}T)^{2}}{\Delta_\text{m}^2\sqrt{4\pi J_0 d^2 \eta}}.
\end{equation}
Moreover, the crossover to $1/f$ noise is pushed to higher frequency. 
In contrast, when $\omega_\text{c}\lesssim \kb T$, only a part of the TLSs relevant for qubit dephasing is affected, which leads to a weaker suppression than $S_\text{white}'$ in Eq.~\eqref{eq:s_low_add}. 
In Fig.~\ref{fig:coherence}(a), we confirm the above analysis by plotting the modified noise spectral densities when the quantum noise is applied. 
The cyan and blue solid curves correspond to the same noise amplitude $d^2\eta$, but with different cutoff frequencies $\omega_\text{c}$. 
Significantly, for $\omega_\text{c}/2\pi=$ 5 GHz, our protocol leads to orders of magnitudes suppression of the white noise amplitude. 
As a result, the qubit pure dephasing time is notably improved, as shown by solid curves in Fig.~\ref{fig:coherence}(b).

\begin{figure*}
    \centering
    \includegraphics[width=0.9\textwidth]{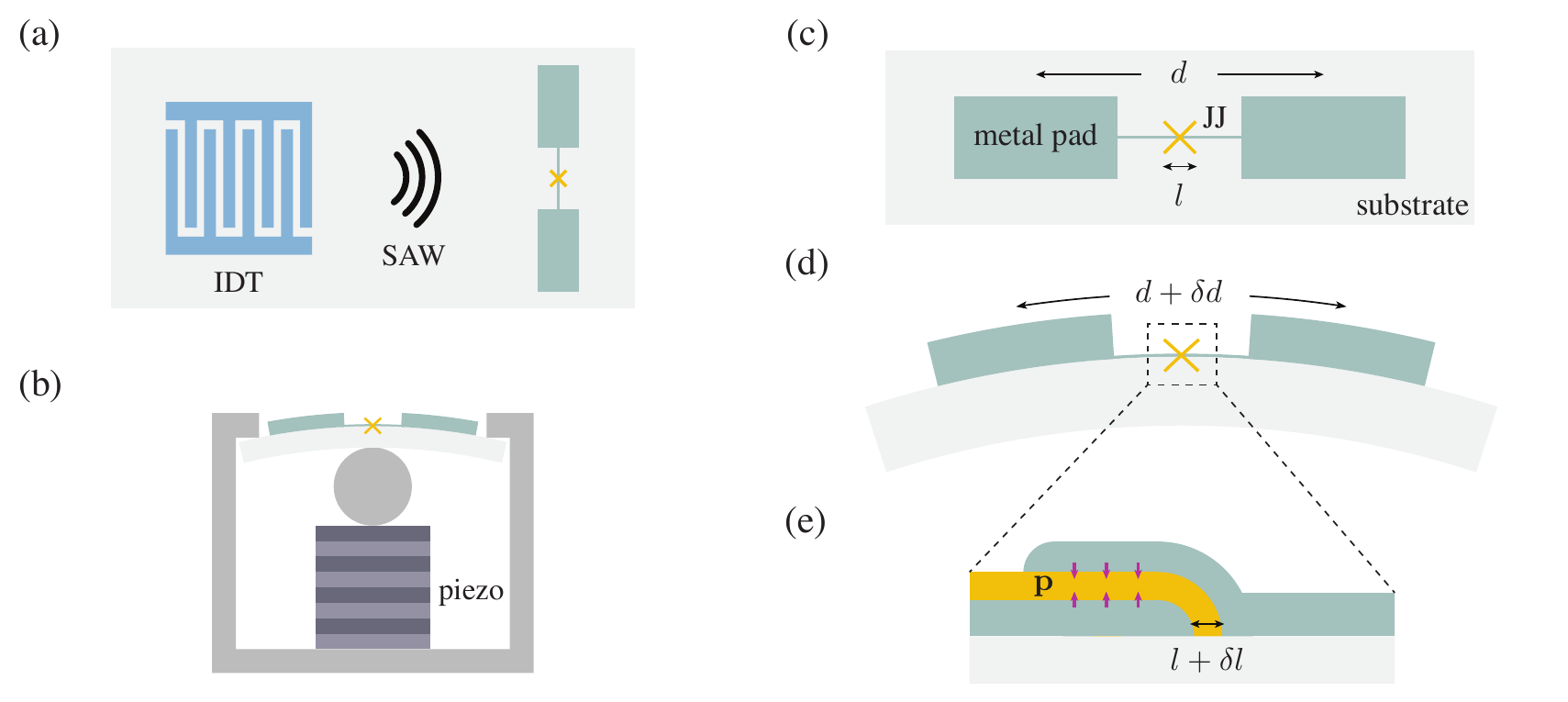}
    \caption{
    Schematic representation of a transmon qubit with applied strain field. 
    (a) An IDT is fabricated on a local piezoelectric thin-film, which generates SAW propagating towards the transmon qubit. 
    (b) A piezo actuator is placed underneath the qubit chip to induce the  deformation. Adapted from Ref.~\cite{Grabovskij2012a}.
    (c) Geometry of a transmon qubit. 
    Green areas denote the metal pads with characteristic distance $d$. 
    The yellow cross symbolize the Josephson junction with width $l$, and the gray background represents the substrate. 
    (d) The qubit chip is deformed in the presence of strain field.
    Specifically, the chip is elongated horizontally, and bent vertically. 
    (e) Zoom-in of the junction region in the presence of deformation. 
    The width of the oxide layer is increased by $\delta l$.
    Purple arrows on the interfaces represent the phonon-induced dipoles $\mathbf{p}$. }
    \label{fig:piezo} 
\end{figure*}

\paragraph{Classical noise}
The modified $S_{n_\text{g}}(\omega)$ with classical noise is shown as dashed curves in Fig.~\ref{fig:coherence}(a).
While the spectra are similar between the cases of quantum and classical noise when $\omega_\text{c}\lesssim \kb T$ (cyan), they differ drastically for $\omega_\text{c}\gg \kb T$ (blue). 
In particular, the modified spectrum for $\omega_\text{c}/2\pi=5$ GHz  is elevated with respect to its quantum counterpart, and is even larger than the one without applied noise (when $\omega/2\pi > 10^{-3}$ Hz). 
This non-monotonic behavior of the modified spectrum regarding $\omega_c$ can be understood as follows.
In thermal equilibrium, only the TLSs with energy smaller than temperature contribute significantly to qubit dephasing.
When $\omega_\text{c}\approx \kb T$, the applied noise selectively depolarizes the same part of TLSs, suppressing the low-frequency noise spectrum.
However, for larger $\omega_\text{c}\gg \kb T$, more TLSs start to contribute to qubit dephasing due to their reduced polarization from the applied classical noise. 
This non-monotonic behavior is also reflected in qubit dephasing time, which first increases and then decreases as a function of the cutoff frequency, as shown by dashed curves in Fig.~\ref{fig:coherence}(b). 

\section{Feasible experimental realizations with negligible spurious coupling to qubits}\label{sec:spurious}

In Secs.~\ref{sec:t1} and~\ref{sec:t2}, we demonstrated in detail how the applied noise modifies the noise spectral densities of an ensemble of TLSs, which results in stabilization of qubit lifetime and improvement in qubit pure dephasing time. 
In practice, the applied noise may also directly couple to the qubit, causing unwanted decoherence. 
However, as will be shown in the following, this is not a concern when the applied noise is in the form of strain field. 
For concreteness, we quantitatively estimate the effects of relevant noise channels on a transmon qubit~\cite{Koch2007,Houck2007}, one of the most popular types of superconducting qubits.
With $E_\text{C}$ and $E_\text{J}$ being its charging and junction energies, the transmon Hamiltonian is 
\begin{equation}
    \hat{H}_\text{q} = 4E_\text{C} (\hat{n}-n_\text{g})^2 - E_\text{J}\cos\hat{\varphi}.
\end{equation}
Here, $\hat{n}$ and $\hat{\varphi}$ are the charge and phase operators of the qubit degree of freedom. The offset charge dependence is manifested in $n_\text{g}$.

Mechanical strain field is a popular and powerful tool to study TLSs. 
For example, strain field has been used to tune the frequencies of TLSs~\cite{Grabovskij2012a,Lisenfeld2015a,Lisenfeld2019,Bilmes2021}, to extract their decoherence properties~\cite{Lisenfeld2016,Brehm2017d}, and to demonstrate spectral hole-burning~\cite{Andersson2021}. 
Experimentally, strain field can be generated in various ways~\cite{OConnell2010,Grabovskij2012a,Gustafsson2014,Teissier2014,Barfuss2015,Barfuss2015,Satzinger2018a,Andersson2021}. 
In Fig.~\ref{fig:piezo}, we sketch two feasible realizations that can be potentially used for the protocols developed in Secs.~\ref{sec:t1} and~\ref{sec:t2}. 
First, as shown in Fig.~\ref{fig:piezo}(a), an interdigital transducer (IDT) is fabricated on a piezoelectric thin film. With applied electrical signals, the IDT generates surface acoustic waves (SAWs) that propagate elastically on the surface of the non-piezoelectric substrate. 
By placing the transmon circuit near the IDT, strain field is induced on those TLSs coupled to the qubit. 
An alternative setup is depicted in Fig.~\ref{fig:piezo}(b), where a stacked piezo actuator is placed under the qubit chip. By applying voltage onto the piezo actuator, it deforms the chip and generates the required strain field. 
We may compare the two realizations by focusing on the frequency and amplitude of the strain field that can be supported, which are two crucial parameters of the proposed protocols. 
In particular, the largest frequency and amplitude used in Secs.~\ref{sec:t1} and~\ref{sec:t2} are on the order of $\omega/2\pi \sim 1\,\text{GHz}$ and $|\mathbf{S}|\sim 10^{-6}$, respectively. 
While the required field amplitude can be achieved in both realizations~\cite{Grabovskij2012a,Andersson2021}, the GHz range frequency is routinely reached in SAWs~\cite{Gustafsson2014,Satzinger2018a,Andersson2021}, but challenging for piezo actuators~\cite{Teissier2014,Barfuss2015}. 

The main advantage of using strain field to address TLS, is that it does not produce stray electromagnetic fields that directly couple to the qubit electric dipole. 
Nevertheless, it does interact with the qubit via other mechanisms. 
In the following, we carefully analyze each of these coupling mechanisms, and confirm that they barely contribute to qubit decoherence. 

\paragraph{Capacitive loss}
The transmon circuit [Fig.~\ref{fig:piezo}(c)] consists of two metal pads connected by a Josephson junction. One consequence of the deformation is that it changes the circuit parameters. 
In particular, the distance $d$ between the two metal pads are modified [Fig.~\ref{fig:piezo}(d)], resulting in a modulation of the capacitance $C\propto d^{-1}$, and thus the charging energy $E_\text{C}=e^2/2C$. 
For small deformation, the perturbed qubit Hamiltonian is 
\begin{equation}
    \delta \hat{H}_\text{cap} = \dfrac{\partial \hat{H}_\text{q}}{\partial E_\text{C}}\dfrac{\partial E_\text{C}}{\partial d} \delta d = 4E_\text{C}(\hat{n}-n_\text{g})^2 \dfrac{\delta d}{d},
\end{equation}
where the deformation relates to strain via $S=\delta d /d$. 

\paragraph{Junction loss}
Besides the distance between metal pads, the thickness of the junction tunnel barrier $l$ also changes in the presence of strain field [Fig.~\ref{fig:piezo}(e)]. 
Since the junction energy has an exponential dependence on the width, $E_\text{J}\propto \exp (-l/l_\text{c})$, the modulation in $E_\text{J}$ is more significant compared with $E_\text{C}$,
\begin{equation}
    \delta \hat{H}_\text{JJ} = \dfrac{\partial \hat{H}_\text{q}}{\partial E_\text{J}}\dfrac{\partial E_\text{J}}{\partial l} \delta l = E_\text{J}\cos\hat{\varphi} \dfrac{\delta l}{l_\text{c}},
\end{equation}
with strain $S=\delta l / l$. For a typical Al/AlO$_x$/Al Josephson junction, the relevant parameters are $l=2$ nm and $l_\text{c}=0.069$ nm~\cite{Gloos2003}. 

\paragraph{Phonon-induced loss}
It is well known that thermal-excited mechanical oscillations in a piezoelectric material produce electric fields via piezoelectric effect. 
The noisy electric fields can then couple to the qubit electric dipole and cause decoherence.
To avoid this loss channel, superconducting qubits are usually fabricated with non-piezoelectric material. 
However, due to surface relaxation~\cite{Ioffe2004,Diniz2020},  electric dipoles can still develop on the interface or surface [purple arrows in Fig.~\ref{fig:piezo}(e)],
\begin{equation}\label{eq:piezo_dipole}
    \mathbf{p} \approx g_\text{I} S t_\text{I} \delta(z)   \hat{z}. 
\end{equation}
Here, ${g}_\text{I}$ is the piezoelectric coefficient, and $t_\text{I}$ is the thickness of the interface. 
The direction of the induced dipole is perpendicular to the interface.
We then apply Eq.~\eqref{eq:piezo_dipole} to the oxide region of the junction, where the qubit mode has maximum electric field amplitude.
For this dielectric/metal (Al$_\text{2}$O$_\text{3}$/Al) interface, the parameters are $g_\text{I}=0.06\,\text{C}/\text{m}^2$ and $t_\text{I}=2.17\,\SI{}{\angstrom}$~\cite{Diniz2020}. 
The phonon-induced dipole interaction is given by
\begin{equation}\label{eq:ph}
    \delta \hat{H}_\text{ph} = \int \mathrm{d}\mathbf{r}\, \mathbf{p}(\mathbf{r})\cdot \hat{\mathbf{E}}_\text{q}(\mathbf{r}) \approx
    \dfrac{4E_\text{C}g_\text{I}t_\text{I}A S}{e l } \hat{n},
\end{equation}
where $A=100\,\textrm{nm}\times100\,\textrm{nm}$ is the typical junction area. 
In deriving the above equation, we approximate the electric field inside the junction to be a uniform field  $|\hat{\mathbf{E}}_\text{q}(\mathbf{r})|\approx\hat{n}4E_\text{C}/(e l)$. 
It should be emphasized here that Eq.~\eqref{eq:ph} overestimates this phonon-induced interaction (and thus the relevant loss), with the following reason.  
The electric dipoles are induced on both sides of the oxide layer, with directions opposite with each other. 
Ideally, the contribution from these two sides should cancel with each other. 
In practice, the cancellation is imperfect due to factors such as surface roughness.

\begin{figure}
    \centering
    \includegraphics[width=0.45\textwidth]{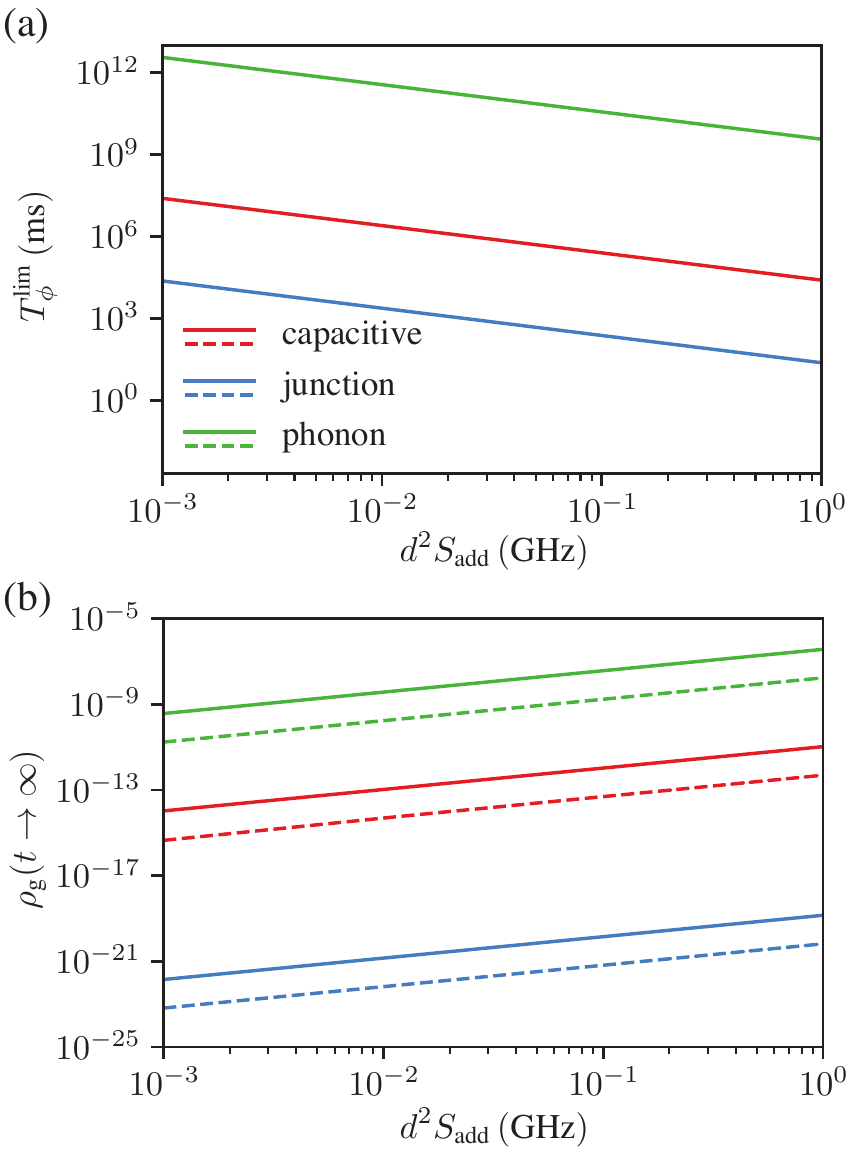}
    \caption{
    Estimation of qubit decoherence from spurious couplings.
    (a) Qubit pure dephasing time for various loss channels, as a function of the applied noise strength.
    (b) Saturated ground state population with the qubit initialized in the excited state.
    Solid and dashed lines represent the cutoff frequencies of 5 GHz and 1 GHz, respectively. [Qubit parameters: $E_\text{C}/h=0.3$ GHz, $E_\text{J}/h=15$ GHz.]}
    \label{fig:spurious} 
\end{figure}

Having introduced the relevant noise channels, we next quantify the effects of those spurious couplings on qubit depolarization and dephasing, respectively. 
For concreteness, we take the applied noise to be white, which corresponds to the case in Sec.~\ref{sec:t1} and also serves as an upper bound for Ohmic noise considered in Sec.~\ref{sec:t2}. 
Since the applied noise is regular at low-frequency, the qubit pure dephasing rate for the noise channel $\lambda$ can be estimated by  
\begin{equation}
    \Gamma_{\phi}^\lambda = \dfrac{1}{2} \big\vert \langle 0 \vert \partial_S\delta\hat{H}_\lambda \vert 0\rangle - \langle 1 \vert \partial_S\delta\hat{H}_\lambda \vert 1\rangle   \big\vert^2 S_\text{add}.
\end{equation}
Note that we have chosen the offset charge to maximize $\Gamma_{\phi}^\lambda$, such that the above expression gives an upper bound for the dephasing rate. 
The predicted limits of qubit pure dephasing time $T^\text{lim}_\phi = 1/\Gamma_{\phi}^\lambda$ for various loss channels are shown in Fig.~\ref{fig:spurious}(a), as a function of the applied noise strength. 
For the largest noise amplitude considered here, all three loss mechanisms pose an upper bound of qubit pure dephasing time above 10 ms. 
Therefore, the decoherence from the spurious coupling is not a concern for the current superconducting qubits.

The depolarization dynamics of the qubit due to additional noise requires careful treatment. 
Here, the cutoff frequency $\omega_\text{c}$ of the applied noise is lower than the qubit frequency.
The Fermi's golden rule would result in zero decay rate of the qubit, i.e., $\Gamma_1^\lambda \propto S_\text{add}(\omega_\text{q}) = 0$, which is incorrect. 
An appropriate description of the qubit depolarization can be derived, for example, following Ref.~\cite{Preskill2015LectureNF}.
Assume the qubit starts in the excited state, its population evolves as
\begin{equation}
    \rho_\text{e}^\lambda(t) = \exp \left[  -S_\text{add} \big\vert \langle 0 \vert \partial_S\delta\hat{H}_\lambda \vert 1\rangle   \big\vert^2 D(t)
    \right],
\end{equation}
with the depolarization function,
\begin{equation}
    D(t) = 2t^2  \int_{-\omega_\text{c}}^{\omega_\text{c}} \dfrac{\mathrm{d}\omega}{2\pi}\,    \text{sinc}^2 \dfrac{(\omega-\omega_\text{q})t}{2}.
\end{equation}
In the limit $\omega_\text{c}\to \infty$, we have $D(t)\to 2t$, which recovers the Markovian result.
However, for a finite $\omega_\text{c}$ considered here, the depolarization function oscillates and saturates on a timescale of $\omega_\text{c}^{-1}$~\footnote{Similar saturation is discussed in the context of qubit dephasing with a super-Ohmic noise with cutoff, for example, in Ref.~\cite{Shnirman2002}.},
\begin{equation}
    D(t\to\infty) \equiv D_\infty = 
    \dfrac{1}{2\pi}\dfrac{8\omega_\text{c}}{\omega_\text{q}^2-\omega_\text{c}^2}.
\end{equation}
Therefore, we choose to characterize qubit depolarization by its ground state population at $t\to\infty$,
\begin{equation*}
    \rho_\text{g}^\lambda(t\to\infty) = 
    1- \exp \left(-S_\text{add} \big\vert \langle 0 \vert \partial_S\hat{H}_\lambda \vert 1\rangle   \big\vert^2 D_\infty
    \right).
\end{equation*}
The results are plotted in Fig.~\ref{fig:spurious}(b) for $\omega_\text{c}=5$ GHz (solid) and 1 GHz (dashed), respectively. 
The largest ground state population in the estimation is still below $10^{-6}$, which confirms the irrelevance of spurious couplings on qubit depolarization. 

For completeness, we briefly comment on the usage of electric field to engineer noise spectrum of TLSs. 
Unlike the strain field, the electric field directly couples to the qubit via electric-dipole interaction,
\begin{equation}
    \delta \hat{H}_\mathbf{E} = \mathbf{d}_\text{q}\cdot \mathbf{E} \,\hat{n},
\end{equation}
where $d_\text{q}=2e L_\text{q}$ is the electric dipole moment of the qubit, and $L_\text{q}$ is its characteristic length. 
For a transmon with $L_\text{q}\sim 10\,\mu\text{m}$~\cite{Koch2007}, its dipole moment is about $d_\text{q}\sim 10^6$ D, six orders of magnitude larger than that of a TLS.  
Even the strength of this spurious interaction could be reduced by aligning the applied electric field to be perpendicular to the qubit chip plane, the residual interaction is still large enough to cause strong qubit decoherence. 
Therefore, while electric field is useful as a tool to dc bias the energy of TLSs, it is not suitable as a method to dynamically engineer the TLSs noise spectrum. 

\section{Conclusions}\label{sec:conclusions}
To summarize, we developed protocols to stabilize and improve qubit coherence times by engineering the relevant noise spectral densities of an ensemble of TLSs.
Specifically, our calculations showed that the high-frequency noise spectrum gets flattened in the presence of applied longitudinal noise on TLSs, leading to an order of magnitude stabilization of qubit lifetime. 
Moreover, we demonstrated that the low-frequency noise amplitude is significantly suppressed via the implementation of transverse noise on TLSs, which results in a longer qubit pure dephasing time. 
Finally, we discussed feasible experimental realizations and confirmed that the spurious coupling from the applied noise does not limit the qubit performance. 
Our results thus suggest a complementary approach to improve qubit coherence, in addition to better qubit design and material investigation.

\begin{acknowledgments}
We thank G. Andersson, S. Chakram, P. Groszkowski, C.R.H. McRae, A. Murthy, J. Mutus, R. Pilipenko, Y.M. Pischalnikov, O.V. Pronitchev, and J.A. Sauls for illuminating discussions.
This material is based upon work supported by the U.S. Department of Energy, Office of Science, National Quantum Information Science Research Centers, Superconducting Quantum Materials and Systems Center (SQMS) under contract number DE-AC02-07CH11359.
\end{acknowledgments}

\appendix

\section{Derivation of central results}\label{app:details}

\subsection{Average qubit lifetime: $\langle \Gamma_1 \rangle_\text{ens}$ in Eq.~\eqref{eq:avg_ens}}
To evaluate the integral in Eq.~\eqref{eq:avg_real}, it is convenient to perform a variable transformation from $(\varepsilon,\Delta)$ to $(\omega_\text{t}, \gamma_2)$. The joint probability distribution of the new variables is 
\begin{equation*}
    P(\omega_\text{t},\gamma_2) = \mathcal{N}(\alpha) \dfrac{\omega_\text{t}}{2(\gamma_2-\gamma_\phi)}
    \left[\omega_\text{t}^2 - \dfrac{\gamma_2-\gamma_\phi}{f(\omega_\text{t})}  \right]^{\tfrac{\alpha - 1}{2}},
\end{equation*}
with function $f(\omega) = \pi J_0\omega\coth(\omega/2\kb T)$. 
In the case of $\alpha=1$, Eq.~\eqref{eq:avg_real} equals to 
\begin{equation*}
    \langle \Gamma_1 \rangle _\text{ens} = \kappa^2 M_1^2 \mathcal{N} \int_{\omega_\text{m}}^{\omega_\text{M}}\dfrac{\mathrm{d}\omega_\text{t}}{\omega_\text{t}f(\omega_\text{t})}
    \int_{\gamma_\text{2,m}(\omega_\text{t})}^{\gamma_\text{2,M}(\omega_\text{t})}\dfrac{\gamma_2\, \mathrm{d}\gamma_2}{\gamma_2^2 + \omega_\delta^2}.
\end{equation*}
The boundaries of the integral are listed as follows: $\omega_\text{m}=\Delta_\text{m}$, $\omega_\text{M}=\sqrt{2}\Delta_\text{M}$, $\gamma_\text{2,m}(\omega_\text{t}) = \Delta_\text{m}^2 f(\omega_\text{t}) + \gamma_\phi$, and $\gamma_\text{2,M}(\omega_\text{t}) = \omega_\text{t}^2 f(\omega_\text{t}) + \gamma_\phi$.
Direct integration of $\gamma_2$ leads to
\begin{equation*}
    \langle \Gamma_1 \rangle _\text{ens} = \kappa^2 M_1^2 \mathcal{N} \int_{\omega_\text{m}}^{\omega_\text{M}}\dfrac{\mathrm{d}\omega_\text{t}}{2\omega_\text{t}f(\omega_\text{t})} 
    \log\left( \dfrac{\gamma_\text{2,M}^2+ \omega_\delta^2}{\gamma_\text{2,m}^2+ \omega_\delta^2}\right).
\end{equation*}
We then approximate the logarithm with a delta function, and integrate over $\omega_\text{t}$,
\begin{align*}
    \langle \Gamma_1 \rangle _\text{ens} &= \kappa^2 M_1^2 \mathcal{N}\pi \int_{\omega_\text{m}}^{\omega_\text{M}}\mathrm{d}\omega_\text{t}\,\dfrac{(\gamma_\text{2,M}-\gamma_\text{2,m})\delta(\omega_\text{t}-\omega)}{\omega_\text{t}f(\omega_\text{t})} 
      \\
   &= \kappa^2 M_1^2 \mathcal{N} \pi \omega_\text{q}. 
\end{align*}
Taking similar steps for $\alpha=0$,  Eq.~\eqref{eq:avg_real} transforms to 
\begin{align*}
    \langle \Gamma_1 \rangle _\text{ens} &= \kappa^2 M_1^2 \mathcal{N} \int_{\omega_\text{m}}^{\omega_\text{M}}\dfrac{\mathrm{d}\omega_\text{t}}{\omega_\text{t}f(\omega_\text{t})} \\
    &\times 
    \int_{\gamma_\text{2,m}(\omega_\text{t})}^{\gamma_\text{2,M}(\omega_\text{t})}\dfrac{\gamma_2 \,\mathrm{d}\gamma_2}{\gamma_2^2 + \omega_\delta^2} 
    \left[ \omega_\text{t}^2 - \dfrac{\gamma_2-\gamma_\phi}{f(\omega_\text{t})}  \right]^{-\tfrac{1}{2}},
\end{align*}
which results in
\begin{equation*}
    \langle \Gamma_1 \rangle _\text{ens} = \kappa^2 M_1^2 \mathcal{N} \int_{\omega_\text{m}}^{\omega_\text{M}}
    \dfrac{2\gamma_\phi\,\mathrm{d}\omega_\text{t}}{\gamma_\phi^2 + \omega_\delta^2} = 2\kappa^2 M_1^2 \mathcal{N}\pi.
\end{equation*}

\subsection{Variance of qubit lifetime: $\text{Var}_\text{ens}(\Gamma_1)$ in Eq.~\eqref{eq:var_ens} and $\langle \text{Var}_\text{spd}(\Gamma_1)\rangle_\text{ens}$ in Eq.~\eqref{eq:var_spd}}
Here, we derive the variance of qubit lifetime over TLS realizations for the case of $\alpha=1$.
The other results can be obtained similarly. 
After variable transformation, Eq.~\eqref{eq:var_ens_org} becomes
\begin{align*}
    \text{Var}_\text{ens}(\Gamma_1) &= \kappa^4 M_1^4 \mathcal{N}
    \int_{\omega_\text{m}}^{\omega_\text{M}}
    \dfrac{2\,\mathrm{d}\omega_\text{t}}{\omega_\text{t}^3f^2(\omega_\text{t})} \\
    &\times
    \int_{\gamma_\text{2,m}(\omega_\text{t})}^{\gamma_\text{2,M}(\omega_\text{t})}
    \dfrac{\gamma_2^2 (\gamma_2-\gamma_\phi)\, \mathrm{d}\gamma_2}{\left(\gamma_2^2 + \omega_\delta^2\right) ^2}.
\end{align*}
Assuming that the TLS pure dephasing rate is much larger than its decay rate, we can employ the approximation,
\begin{equation*}
    \gamma_2^2 (\gamma_2-\gamma_\phi)/\left(\gamma_2^2 + \omega_\delta^2\right) ^2 \approx 
    \gamma_\phi^2 (\gamma_2-\gamma_\phi)/(\gamma_\phi^2 + \omega_\delta^2) ^2.
\end{equation*}
With the boundaries of $\omega_\text{t}$ being extended to $(0,\infty)$, we derive the results shown in Eq.~\eqref{eq:var_ens},
\begin{align*}
    \text{Var}_\text{ens}(\Gamma_1) &= \kappa^4 M_1^4 \mathcal{N}^2
    \int_{\omega_\text{m}}^{\omega_\text{M}}
    \dfrac{2\gamma_\phi^2\omega_\text{t} \,\mathrm{d}\omega_\text{t}}{(\gamma_\phi^2 + \omega_\delta^2) ^2}\\
    & = \kappa^4 M_1^4 \mathcal{N}\pi\omega_\text{q}/2\gamma_\phi.
\end{align*}

\subsection{Low frequency noise spectral density of an ensemble of TLSs: $S_\text{low}(\omega)$ in Eq.~\eqref{eq:s_low} and $S_\text{white}'$ in Eq.~\eqref{eq:s_low_add}}
With variable transformation from $(\varepsilon,\Delta)$ to $(\omega_\text{t},\gamma_1)$, the low-frequency noise $S_\text{low}(\omega)$ is given by
\begin{equation*}
    S_\text{low}(\omega)  =N_\text{TLS} \int_{\omega_\textrm{m}}^{\omega_\textrm{M}} \mathrm{d}\omega_\textrm{t}  \int_{\gamma_\text{1,m}}^{\gamma_\text{1,M}} \mathrm{d}\gamma_1 P(\gamma_1, \omega_\text{t} ) \cos^2\theta s_{zz}(\omega).
\end{equation*}
Here, the joint probability distribution is 
\begin{equation*}
     P(\gamma_1, \omega_\text{t} )=  \mathcal{N} \omega_\text{t}/2\gamma_1,
\end{equation*}
with $\mathcal{N}$ the normalization factor.
The boundaries of the integral are $\gamma_\text{1,m}(\omega_\text{t}) = 2\Delta_\text{m}^2 f(\omega_\text{t})$, and $\gamma_\text{1,M}(\omega_\text{t}) = 2 \omega_\text{t}^2 f(\omega_\text{t})$. 
With $\cos^2\theta = 1 - \gamma_1/\gamma_\text{1,M}$,
we have,
\begin{widetext}
\begin{align*}
    S_\text{low}(\omega) =& N_\text{TLS} \mathcal{N} \int_{\omega_\textrm{m}}^{\omega_\textrm{M}} \mathrm{d}\omega_\textrm{t} \,\omega_\text{t} \int_{\gamma_\text{1,m}}^{\gamma_\text{1,M}} \mathrm{d}\gamma_1
     \dfrac{1-\langle \hat{\sigma}_z \rangle_\text{eq}^2}{\omega^2  + \gamma_1^2} \left(1 - \dfrac{\gamma_1}{\gamma_\text{1,M}}\right) \\
    =&  N_\text{TLS}\mathcal{N} \int_{\omega_\textrm{m}}^{\omega_\textrm{M}} \mathrm{d}\omega_\textrm{t}\, \omega_\textrm{t}\left\{ 
    \dfrac{1}{\omega}\left[\tan^{-1}\left(\frac{\gamma_\textrm{1,M}}{\omega}\right) - \tan^{-1}\left( \dfrac{\gamma_\textrm{1,m}}{\omega}\right)\right]
    - \dfrac{1}{2\gamma_\textrm{1,M}} \log (\dfrac{\gamma_\textrm{1,M}^2 + \omega^2}{\gamma_\textrm{1,m}^2 + \omega^2})
    \right\}(1-\langle \hat{\sigma}_z \rangle_\text{eq}^2).
\end{align*}
\end{widetext}
For $\gamma_\textrm{m} \ll \omega \ll \gamma_\textrm{M}$, the spectral density has a $1/f$ behavior,
\begin{align*}
    S_\text{low}(\omega) \approx S_{1/f}(\omega) &=
\dfrac{N_\text{TLS}\mathcal{N} \pi}{2\omega}\int_{\omega_\textrm{m}}^{\omega_\textrm{M}} \mathrm{d}\omega_\textrm{t}\, 
\omega_\textrm{t}(1-\langle \hat{\sigma}_z \rangle_\text{eq}^2) \\
& =2\log(2)\pi N_\text{TLS}\mathcal{N}(\kb T)^2/\omega. 
\end{align*}
While for $\omega \ll \gamma_\textrm{m} \ll \gamma_\textrm{M}$, the spectral density exhibits a white noise behavior,
\begin{align*}
    S_\text{low}(\omega) \approx S_\text{white}(\omega) &=
N_\text{TLS}\mathcal{N}\int_{\omega_\textrm{m}}^{\omega_\textrm{M}} \mathrm{d}\omega_\textrm{t}\, \omega_\textrm{t} \,  (1-\langle \hat{\sigma}_z \rangle_\text{eq}^2)/\gamma_\text{1,m} \\
& =\dfrac{N_\text{TLS}\mathcal{N}\kb T}{\Delta_\text{m}^2 2\pi J_0}.
\end{align*}

In the presence of applied quantum noise, the depolarization rate of the TLS is modified according to Eq.~\eqref{eq:add_g1}. 
As a result, the lower boundary of the integral becomes,
\begin{equation*}
    \gamma_\text{1,m}'(\omega,\alpha) = 2\Delta_\text{m}^2 f(\omega_\text{t}) + 2d^2 \eta \cos{\alpha}^2 \Delta_\text{m}^2/\omega_\text{t}. 
\end{equation*}
Reevaluating the white noise part results in 
\begin{align*}
    S_\text{white}'(\omega) &=
    \dfrac{2N_\text{TLS}\mathcal{N}}{\pi}\int_0^{\frac{\pi}{2}}\mathrm{d}\alpha\int_{\omega_\textrm{m}}^{\omega_\textrm{M}} \mathrm{d}\omega_\textrm{t}\, \omega_\textrm{t} \, \dfrac{1-\langle \hat{\sigma}_z \rangle_\text{eq}^2}{\gamma_\text{1,m}'}  \\
    &\approx N_\text{TLS}\mathcal{N}\dfrac{0.58(2k_\text{B}T)^{2}}{\Delta_\text{m}^2\sqrt{4\pi J_0 d^2 \eta}}.
\end{align*}
Note that an additional integration is required to average over the angle $\alpha$ between the applied field and the corresponding TLS dipole, which is assumed to have a uniform distribution.
In the above derivation, we also make the approximation that $\gamma_{1,\text{m}}'$ is dominated by the applied noise, and extend the integral over $\omega_\text{t}$ to $(0,\infty)$.

\bibliography{main.bib}

\end{document}